\documentclass[journal,twoside,web,hyperref={colorlinks =true,linkcolor = blue}, xcolor=dvipsnames]{ieeecolor}
\pdfoutput=1
\usepackage{jsen}
\usepackage{cite}
\usepackage{amsmath,amssymb,amsfonts}
\usepackage{algorithmic}
\usepackage{graphicx}
\usepackage{textcomp}
\usepackage{wrapfig}
\usepackage{bm}
\usepackage[colorlinks,bookmarksopen,bookmarksnumbered,citecolor=blue, linkcolor=blue, urlcolor=blue]{hyperref}
\usepackage{lineno,hyperref}
\usepackage{array}
\usepackage{subfigure}
\usepackage{orcidlink}
\usepackage{amssymb}
\usepackage{xcolor}
\usepackage{ulem}
\def\BibTeX{{\rm B\kern-.05em{\sc i\kern-.025em b}\kern-.08em
    T\kern-.1667em\lower.7ex\hbox{E}\kern-.125emX}}
\definecolor{abstractbg}{rgb}{0.89804,0.94510,0.83137}
\begin{document}
\title{Adaptive radar detection of subspace-based distributed target in power heterogeneous clutter}
\author{Daipeng Xiao\orcidlink{0009-0002-6431-6082}, Weijian Liu\orcidlink{0000-0002-0330-8073}, \IEEEmembership{Senior Member, IEEE}, Jun Liu\orcidlink{0000-0002-7193-0622}, \IEEEmembership{Senior Member, IEEE}, Lingyan Dai, Xueli Fang, and Jianjun Ge\orcidlink{0009-0002-6568-109X}
\thanks{This work was supported in part by the National Natural Science Foundation of China, under Grant 62071482. \textit{(Corresponding author: Weijian Liu)}}
\thanks{D. Xiao, W. Liu and L. Dai are with Wuhan Electronic Information Institute, Wuhan 430019, China. (e-mails: daipxiao@163.com; liuvjian@163.com; daily\_076@163.com).}
\thanks{J. Liu is with the Department of Electronic Engineering and Information Science, University of Science and Technology of China, Hefei 230027, China
	(e-mail: junliu@ustc.edu.cn).}
\thanks{X. Fang is with the Key Laboratory of Complex Aviation Simulation System, Beijing 100076, China (e-mail: fangxueli@163.com).}
\thanks{J. Ge is with the Information Science Academy, China Electronics Technology Group Corporation, Beijing 100041, China (e-mail: geradarnet@163.com).}}
\IEEEtitleabstractindextext{%
\fcolorbox{abstractbg}{abstractbg}{%
\begin{minipage}{\textwidth}%
\begin{wrapfigure}[12]{r}{3in}%
\includegraphics[width=3in]{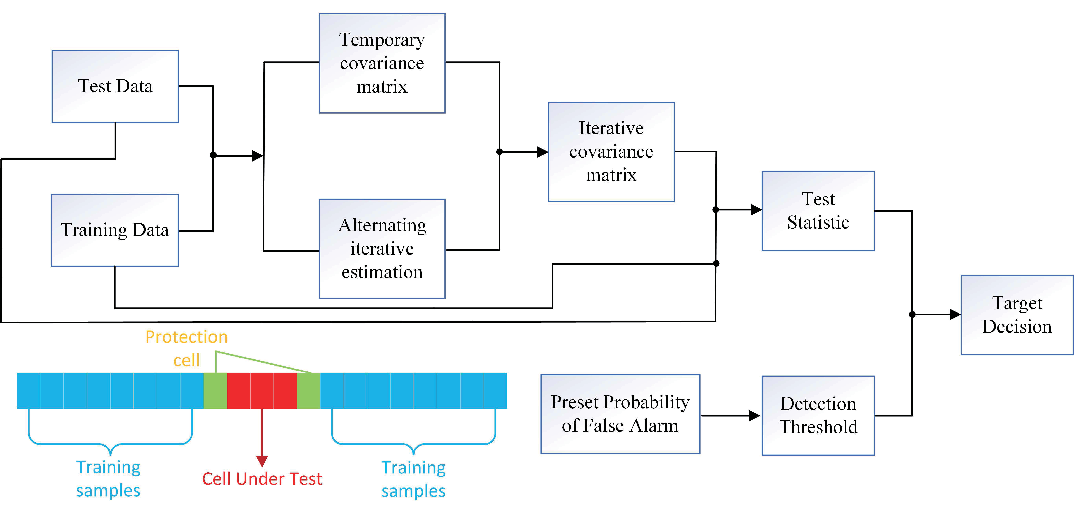}%
\end{wrapfigure}%
\begin{abstract}
This paper investigates the problem of adaptive detection of distributed targets in power heterogeneous clutter. In the considered scenario, all the data share the identical structure of clutter covariance matrix, but with varying and unknown power mismatches. To address this problem, we iteratively estimate all the unknowns, including the coordinate matrix of the target, the clutter covariance matrix, and the corresponding power mismatches, and propose three detectors based on the generalized likelihood ratio test (GLRT), Rao and the Wald tests. The results from simulated and real data both illustrate that the detectors based on GLRT and Rao test have higher probabilities of detection (PDs) than the existing competitors. Among them, the Rao test-based detector exhibits the best overall detection performance. We also analyze the impact of the target extended dimensions, the signal subspace dimensions, and the number of training samples on the detection performance. Furthermore, simulation experiments also demonstrate that the proposed detectors have a constant false alarm rate (CFAR) property for the structure of clutter covariance matrix.
\end{abstract}

\end{minipage}}}

\maketitle

\section{Introduction}
\label{sec:introduction}
\IEEEPARstart{S}{ince} Kelly proposed an adaptive detector based on the generalized likelihood ratio test (GLRT) in 1986 \cite{Kelly86}, adaptive detection technology has undergone more than 30 years of development and is widely used in fields of radar \cite{ColucciaFascista24}, sonar \cite{KlausnerAzimiSadjadi19}, communications \cite{Ghorbel15IESensor}, and so on. Various adaptive detectors have been proposed for different problems and according to different detector design criteria \cite{LiuLiu22SCIS}.

As radar technology progresses, the performance of radar systems has continuously enhanced \cite{Sen14,Zaimbashi17IESensor}, resulting in a consistent improvement in the radar's resolution as well \cite{BluntGerlach07}. It is widely known that a target is often extended in the range domain in high-resolution radar (HRR). Hence, the target usually occupies several radar range resolution cells, and it is commonly acknowledged as a distributed target \cite{Xu2016adaptive,LinChen15}.
Over the course of the last few years, the problem of adaptive detection for distributed targets has been widely studied \cite{BandieraDeMaio07a}, \cite{LiuZhang14}, \cite{CuiWang23JSEE}. Given the supposition that all echoes come from the same direction, Conte \textit{et al.} derived the GLRT in homogeneous and partially homogeneous environments \cite{ConteDeMaio01}, and the corresponding Rao and the Wald tests were obtained in \cite{ShuaiKong12} and \cite{Hao2012}, respectively. In \cite{ConteDeMaio03}, the detection problem of distributed target with unknown signal steering vector had been investigated. Considering that the number of training samples may be scarce in the real environment, Wang \textit{et al.} incorporated the persymmetric structure into the design of detector to propose adaptive detectors in partially homogeneous Gaussian environments \cite{WangLi16SP}, and Gao \textit{et al.} utilized the Bayesian approach, modeling the unknown covariance matrix as an inverse Wishart distribution to propose detectors based on GLRT \cite{GaoLiao17RSN}. The authors in \cite{Li_Liu17CSSP} introduced a thorough examination of the statistical properties of the detectors employed for distributed target detection, with the condition that covariance matrix is known. In addition, taking into account that the number of unknown parameters may be unknown in detection, based on the Kullback-Leibler (KL) information criterion, Faro \textit{et al.} addressed the multiple hypothesis testing problem by considering the Kullback-Leibler Divergence (KLD) between the candidate distribution probability density function (PDF) and the actual PDF, and applied it to subspace target detection in the presence of multiple alternative hypotheses in \cite{FaroGiunta20}. Then, in \cite{Addabbo2021adaptive}, Addabbo \textit{et al.} provided a detailed design scheme for adaptive detection based on the KL information criterion under multiple hypothesis testing, and proved the constant false alarm rate (CFAR) property of the proposed detectors.

The detection problems mentioned above are all based on homogeneous Gaussian or partially homogeneous Gaussian models. The former model means that the clutter in all range bins can be considered as having the same statistical characteristics that conform to a Gaussian distribution. In the latter model, in addition to satisfying the aforementioned conditions, there is also a definite but unknown power mismatch between the cell under tested (CUT) and training samples. But real scenarios may not conform to these assumptions and could be non-homogeneous and non-Gaussian \cite{HaoOrlando12a}, \cite{LiuMassaro20}, \cite{Xiao2024}, \cite{HE2011Two}. Through the study of scattering theory and the analysis of a large amount of measured clutter data \cite{ConteDeMaio05, GrecoGini06}, some scholars have employed compound Gaussian clutter model to describe the statistical characteristics of the clutter \cite{GiniGreco02b}. This model is defined as the result of the multiplication of slowly varying texture component and rapidly varying speckle component \cite{Mezache15AES}, and the amplitude distribution of clutter varies depending on the probability distributions of the texture component. Within this group, frequently used clutter amplitude distribution models include K \cite{Shui2016TAES}, Pareto \cite{Weinberg2013b} and CG-IG distributions \cite{Ollila12SPL}, etc. The detection problem of distributed targets in compound Gaussian clutter has been studied in a significant amount of references \cite{Guo24PersymmetricSP}, \cite{JiangWu23}, \cite{Xu22AES}.
For example, the distributed target detectors based on GLRT, as well as Rao and the Wald tests in compound Gaussian clutter were proposed in \cite{ShuaiKong10a} and \cite{ConteDeMaio03a}. In addition, Xu \textit{et al.} incorporated \textit{a priori} distribution of clutter into the design of detectors and also considered the extension of the target in both range and Doppler domains \cite{XuShi18}. In the construction of detectors within compound Gaussian clutter, it usually needs to assign \textit{a priori} distribution of texture component, which can lead to the adaptive detectors that outperform the detectors not accounting for the texture distribution in terms of detection performance \cite{Sangston99AES, SangstonGini12}. 
The actual texture component may have a statistical distribution different from the assumed distribution, and even the clutter does not necessarily obey the compound Gaussian distribution.
In such cases, detectors designed based on the predefined statistical distribution of the texture component will suffer from the performance degradation.

To address the above issues, it is assumed in \cite{Coluccia2022} that the data, including CUT and training samples, all share identical structure of clutter covariance matrix. However, different power mismatches are present among the training samples, which are modeled as non-random and non-negative values. Direct computation of the maximum likelihood estimation (MLE) for all values of power mismatches fails to produce a closed-form solution. Consequently, Coluccia \textit{et al.} have utilized an alternating iterative estimation approach, based on the GLRT criterion to propose an adaptive detector \cite{Coluccia2022}. Experiments have demonstrated that the proposed detector outperforms adaptive normalized matched filter (ANMF) \cite{Rangaswamy05}, and it has been proved that when using specific predefined values, the detector maintains a CFAR property for the structure of clutter covariance matrix.

It should be noted that the adaptive detector is designed solely based on GLRT criterion in \cite{Coluccia2022}, and the detector is only suitable for the model of rank-one signals specifically for a point target. However, the target may be a distributed target, and the signal can be characterized as a subspace signal. Additionally, when it comes to signal detection in complex real scenarios, the detection environment and signal parameters are often unknown, rendering the existence of a uniformly most powerful (UMP) test impossible. Besides GLRT criterion, other frequently used detector design criteria are Rao and the Wald tests.

In this paper, we delve deeper into the detection problem originally considered in \cite{Coluccia2022}, by generalizing rank-one point target to subspace-based distributed target and using GLRT, Rao and the Wald tests, combined with the alternating iterative estimation to propose three detectors, termed the AIE-GLRT, AIE-Rao and AIE-Wald. Remarkably, experiments on simulated and real data both demonstrate that the PDs of AIE-Rao and AIE-GLRT are higher than those of the existing competitors, with the AIE-Rao having the best performance. Moreover, simulation experiments also illustrate that all detectors have a CFAR property for clutter covariance matrix structure. 

Notation: The symbols ${( \cdot )^ * }$, ${( \cdot )^{\rm{T}}}$, ${( \cdot )^{ - 1}}$, ${( \cdot )^{\rm{H}}}$, ${( \cdot )^ - }$, $\partial ( \cdot )$, $\text{tr}\left( \cdot  \right)$, $\text{vec}\left( \cdot  \right)$, $| \cdot |$, ${{\left\| \cdot  \right\|}_{\text{F}}}$ and $\otimes $ represent conjugate, transpose, inverse, conjugate transpose, generalized inverse, partial derivative, trace, vectorization, determinant, Frobenius norm and Kronecker product, respectively.

\section{Problem Formulation}
The data to be detected are represented by an $N\times K$ dimensional matrix $\mathbf{Z}$, where $N$ is the number of system channels, and $K$ denotes the count of range bins that a distributed target occupies. Under hypothesis ${{\text{H}}_{0}}$, $\mathbf{Z}$ only contains the clutter components $\mathbf{C}$, where each column of $\mathbf{C}$ is independently and identically distributed (IID). The $k$-th column of $\mathbf{C}$ is represented by ${{\mathbf{c}}_{k}}$, which follows a complex Gaussian distribution characterized by a mean of zero and a covariance matrix of $\mathbf{R}$, expressed as ${{\mathbf{c}}_{k}}\sim\mathcal{C}{{\mathcal{N}}_{N}}(\mathbf{0},\mathbf{R})$. Under hypothesis ${{\text{H}}_{1}}$, $\mathbf{Z}$ includes the clutter components $\mathbf{C}$ and the signal components $\mathbf{S}$, expressed as $\mathbf{S}=\mathbf{H\Psi }$, where an $N\times p$ dimensional matrix $\mathbf{H}$ spans the signal subspace and a $p\times K$ dimensional matrix $\mathbf{\Psi }$ represents the signal coordinates.

It is widely known that the covariance matrix $\mathbf{R}$ is usually unknown in the real environment. In order to estimate $\mathbf{R}$, it is presumed that there are the number of $L$ training samples, denoted as ${{\mathbf{z}}_{l}},\text{ }l=1,2,\cdots ,L$, where ${{\mathbf{z}}_{l}}$ only includes clutter components ${{\mathbf{c}}_{l}}$. Hence, the problem of detecting distributed targets can be modeled as a binary hypothesis test
\begin{equation}\label{1}
	\left\{ \begin{array}{*{35}{l}}
		{{\text{H}}_{0}}:\text{ }\mathbf{Z}=\mathbf{C}\text{,  }{{\mathbf{Z}}_{L}}={{\mathbf{C}}_{L}}    \\
		{{\text{H}}_{1}}:\text{ }\mathbf{Z}=\mathbf{H\Psi }+\mathbf{C}\text{,  }{{\mathbf{Z}}_{L}}={{\mathbf{C}}_{L}}    \\
	\end{array} \right.
\end{equation}
where ${{\mathbf{Z}}_{L}}=\left[ {{\mathbf{z}}_{1}},{{\mathbf{z}}_{2}},\cdots ,{{\mathbf{z}}_{L}} \right]$, ${{\mathbf{C}}_{L}}=\left[ {{\mathbf{c}}_{1}},{{\mathbf{c}}_{2}},\cdots {{\mathbf{c}}_{L}} \right]$. Suppose ${{\mathbf{c}}_{l}}$ follow a complex Gaussian distribution characterized by a mean of zero and a covariance matrix of ${{\mathbf{R}}_{l}}$, abbreviated as ${{\mathbf{c}}_{l}}\sim\mathcal{C}{{\mathcal{N}}_{N}}(\mathbf{0},{{\mathbf{R}}_{l}})$. In the context of detection in this paper, we can model ${{\mathbf{R}}_{l}}$ as  ${{\mathbf{R}}_{l}}={{\sigma }_{l}}\mathbf{R}\text{,  }l=1,2,\cdots ,L$, where ${{\sigma }_{l}}\text{,  }l=1,2,\cdots ,L$ are the unknown parameters, denoting the varying power mismatches among the data.

\section{Detector Design}
Under hypotheses ${{\text{H}}_{1}}$ and ${{\text{H}}_{0}}$, the jointly PDFs of CUT and training samples can be represented as
\begin{equation}\label{x3}
	\begin{aligned}
		& {{f}_{1}}(\mathbf{Z},{{\mathbf{Z}}_{L}};\mathbf{\Psi },\mathbf{R},\bm{\sigma })=\frac{1}{{{\pi }^{N(L+K)}}\prod\limits_{l=1}^{L}{\sigma _{l}^{N}}|\mathbf{R}{{|}^{L+K}}} \\ 
		& \text{                          }\times \text{etr}\left[ -{{\left( \mathbf{Z}-\mathbf{H\Psi } \right)}^{\text{H}}}{{\mathbf{R}}^{-1}}\left( \mathbf{Z}-\mathbf{H\Psi } \right)-\sum\limits_{l=1}^{L}{\frac{\mathbf{z}_{l}^{\text{H}}{{\mathbf{R}}^{-1}}{{\mathbf{z}}_{l}}}{{{\sigma }_{l}}}} \right] \\ 
	\end{aligned}
\end{equation}
and
\begin{equation}\label{x4}
\begin{aligned}
	{{f}_{0}}(\mathbf{Z},{{\mathbf{Z}}_{L}};\mathbf{R},\bm{\sigma })=&\frac{1}{{{\pi }^{N(L+K)}}\prod\limits_{l=1}^{L}{\sigma _{l}^{N}}|\mathbf{R}{{|}^{L+K}}}\\
	&\times \text{etr}\left( -{{\mathbf{Z}}^{\text{H}}}{{\mathbf{R}}^{-1}}\mathbf{Z}-\sum\limits_{l=1}^{L}{\frac{\mathbf{z}_{l}^{\text{H}}{{\mathbf{R}}^{-1}}{{\mathbf{z}}_{l}}}{{{\sigma }_{l}}}} \right)\\
\end{aligned}
\end{equation}
respectively, where $\bm{\sigma }={{\left[ {{\sigma }_{1}},{{\sigma }_{2}},\cdots ,{{\sigma }_{L}} \right]}^{\text{T}}}$ represents a column vector consisting of the amount of power mismatches.

\subsection{GLRT}
The GLRT can be represented as
\begin{equation}\label{5}
	\frac{{{\max }_{\mathbf{\Psi },\mathbf{R},\bm{\sigma }}}{{f}_{1}}(\mathbf{Z},{{\mathbf{Z}}_{L}};\mathbf{\Psi },\mathbf{R},\bm{\sigma })}{{{\max }_{\mathbf{R},\bm{\sigma }}}{{f}_{0}}(\mathbf{Z},{{\mathbf{Z}}_{L}};\mathbf{R},\bm{\sigma })}\text{ }\underset{{{\text{H}}_{0}}}{\overset{{{\text{H}}_{1}}}{\mathop{\gtrless }}}\,\text{ }\eta 
\end{equation} 
where $\eta $ denotes the detection threshold set based on the pre-designed probability of false alarm (PFA).

Under hypothesis $\text{H}_{1}$, by setting the partial derivative of \eqref{x3} with respect to (w.r.t.) $\mathbf{R}$ and letting it be zero, we obtain the MLE of $\mathbf{R}$ for the given $\mathbf{\Psi}$ and $\bm{\sigma}$ as
\begin{equation}\label{x5}
	{{\mathbf{\hat{R}}}_{1}}=\frac{1}{L+K}\left[ \left( \mathbf{Z}-\mathbf{H\Psi } \right){{\left( \mathbf{Z}-\mathbf{H\Psi } \right)}^{\text{H}}}+\sum\limits_{l=1}^{L}{\frac{{{\mathbf{z}}_{l}}\mathbf{z}_{l}^{\text{H}}}{{{\sigma }_{l}}}} \right]
\end{equation}
In a similar method, we derive the MLE of $\mathbf{R}$ under hypothesis $\text{H}_{0}$ for the given $\bm{\sigma}$ as
\begin{equation}\label{x6}
	{{\mathbf{\hat{R}}}_{0}}=\frac{1}{L+K}\left( \mathbf{Z}{{\mathbf{Z}}^{\text{H}}}+\sum\limits_{l=1}^{L}{\frac{{{\mathbf{z}}_{l}}\mathbf{z}_{l}^{\text{H}}}{{{\sigma }_{l}}}} \right)
\end{equation}
Substituting \eqref{x5} and \eqref{x6} into \eqref{x3} and \eqref{x4} results in 
\begin{equation}\label{x7}
	{{f}_{1}}(\mathbf{Z},{{\mathbf{Z}}_{L}};\mathbf{\Psi },{{\mathbf{\hat{R}}}_{1}},\bm{\sigma })=\frac{{{\left( \frac{L+K}{e\pi } \right)}^{N(L+K)}}}{\prod\limits_{l=1}^{L}{\sigma _{l}^{N}}{{\left. \left| ({{\mathbf{Z}}_{1}}\mathbf{Z}_{1}^{\text{H}}+\sum\limits_{l=1}^{L}{\frac{{{\mathbf{z}}_{l}}\mathbf{z}_{l}^{\text{H}}}{{{\sigma }_{l}}}} \right. \right|}^{(L+K)}}}
\end{equation}
and
\begin{equation}\label{x8}
	{{f}_{0}}(\mathbf{Z},{{\mathbf{Z}}_{L}};{{\mathbf{\hat{R}}}_{0}},\bm{\sigma })=\frac{{{\left( \frac{L+K}{e\pi } \right)}^{N(L+K)}}}{\prod\limits_{l=1}^{L}{\sigma _{l}^{N}}{{\left. \left| ({{\mathbf{Z}}}\mathbf{Z}^{\text{H}}+\sum\limits_{l=1}^{L}{\frac{{{\mathbf{z}}_{l}}\mathbf{z}_{l}^{\text{H}}}{{{\sigma }_{l}}}} \right. \right|}^{(L+K)}}}
\end{equation}
respectively, where  ${{\mathbf{Z}}_{1}}=\mathbf{Z}-\mathbf{H\Psi }$.

Direct derivation of MLE for $\mathbf{\Psi}$ and $\bm{\sigma}$ does not seem to yield a closed-form solution. As a result, we consider the approach of alternating iterative estimation to solve the problem. For convenience, we denote the predefined value and the ultimate estimation of $\bm{\sigma}$ under $\text{H}_{1}$ as $\hat{\sigma }_{l}^{1,(0)}$, and $\hat{\sigma }_{l}^{1,({{n}_{\max }})}$, respectively. Similarly, $\hat{\sigma }_{l}^{0,(0)}$ and $\hat{\sigma }_{l}^{0,({{n}_{\max }})}$ symbolize the predefined value and the ultimate estimation of $\bm{\sigma}$ under $\text{H}_{0}$. Moreover, ${{\mathbf{\hat{\Psi }}}^{({{n}_{\max }})}}$ denotes the ultimate estimation of $\mathbf{\Psi}$.

Then, to derive the ultimate estimation for $\mathbf{\Psi}$ through the iterative process, we introduce
\begin{equation}\label{x9}
	f(\mathbf{\Psi })=\left. \left| (\mathbf{Z}-\mathbf{H\Psi }){{(\mathbf{Z}-\mathbf{H\Psi })}^{\text{H}}}+{{\mathbf{\Omega }}^{(n)}} \right. \right|
\end{equation}
where ${{\mathbf{\Omega }}^{(n)}}=\sum\limits_{l=1}^{L}{\frac{{{\mathbf{z}}_{l}}\mathbf{z}_{l}^{\text{H}}}{\hat{\sigma }_{l}^{1,(n)}}}$, the superscript $n$ indicates the $n$-th iterative estimate. According to the matrix determinant lemma
\begin{equation}\label{x10}
	\left| \left. \mathbf{A+BC} \right| \right.=\left| \left. \mathbf{A} \right| \right.\cdot \left| \left. \mathbf{I}+\mathbf{C}{{\mathbf{A}}^{-1}}\mathbf{B} \right| \right.
\end{equation}
with $\mathbf{A}$, $\mathbf{B}$ and $\mathbf{C}$ being matrices of compatible dimensions. Then, we can transform \eqref{x9} as
\begin{equation}\label{x11}
	f(\mathbf{\Psi })=\left| \left. {{\mathbf{\Omega }}^{(n)}} \right| \right.\cdot \left| \left. \mathbf{I}+{{\left( \mathbf{Z}-\mathbf{H\Psi } \right)}^{\text{H}}}{{\left[ {{\mathbf{\Omega }}^{(n)}} \right]}^{-1}}\left( \mathbf{Z}-\mathbf{H\Psi } \right) \right| \right.
\end{equation}
Taking the partial derivative of \eqref{x11} w.r.t. $\mathbf{\Psi}$ to be zero, we have
\begin{equation}\label{x12}
	{{\mathbf{\hat{\Psi }}}^{(n+1)}}={{\left\{ {{\mathbf{H}}^{\text{H}}}{{\left[ {{\mathbf{\Omega }}^{(n)}} \right]}^{-1}}\mathbf{H} \right\}}^{-1}}{{\mathbf{H}}^{\text{H}}}{{\left[ {{\mathbf{\Omega }}^{(n)}} \right]}^{-1}}\mathbf{Z}
\end{equation}

To enhance the clarity of the paper, we have placed the derivation of the iterative estimates of $\bm{\sigma}$ under hypotheses ${{\text{H}}_{1}}$ and ${{\text{H}}_{0}}$ in the Appendix.

 Based on the above derivation and the Appendix, we begin the iteration with the predefined value of $\bm{\sigma}$ under hypothesis $\text{H}_{1}$, denoted as $\hat{\sigma} _{l}^{1,(0)}\text{,  }l=1,2,\cdots,L$. Then we can obtain the $\hat{\mathbf{\Psi}}^{(1)}$ by \eqref{x12}, and substituting $\hat{\mathbf{\Psi}}^{(1)}$ into \eqref{x16} yields $\hat{\sigma} _{l}^{1,(1)}\text{,  }l=1,2,\cdots ,L$. This iterative procedure will terminate once the maximum number of iterations is attained, and the rationale for setting this value will be discussed in next section. Then we can obtain $\hat{\sigma} _{l}^{1,(n_{max})}\text{,  }l=1,2,\cdots ,L$ and $\hat{\mathbf{\Psi}}^{(n_{max})}$ under hypothesis $\text{H}_{1}$.
 Similarly, we can obtain $\hat{\sigma} _{l}^{0,(n_{max})}\text{,  }l=1,2,\cdots ,L$.

Finally, we substitute the ultimate estimations of $\bm{\sigma}$ and $\mathbf{\Psi}$ into \eqref{x7} and \eqref{x8}, and then substituting them into \eqref{5} yields the AIE-GLRT as
\begin{equation}\label{9}
	{{t}_{\text{AIE-GLRT}}}=\frac{{{\left[ \frac{\prod\nolimits_{l=1}^{L}{\hat{\sigma }_{l}^{0,({{n}_{\max }})}}}{\prod\nolimits_{l=1}^{L}{\hat{\sigma }_{l}^{1,({{n}_{\max }})}}} \right]}^{\frac{N}{L+K}}}\left| \left. \mathbf{Z}{{\mathbf{Z}}^{\text{H}}}+\sum\limits_{l=1}^{L}{\frac{{{\mathbf{z}}_{l}}\mathbf{z}_{l}^{\text{H}}}{\hat{\sigma }_{l}^{0,({{n}_{\max }})}}} \right| \right.}{\left| \left. \left( \mathbf{Z}-\mathbf{H}{{{\mathbf{\hat{\Psi }}}}^{({{n}_{\max }})}} \right){{\left( \mathbf{Z}-\mathbf{H}{{{\mathbf{\hat{\Psi }}}}^{({{n}_{\max }})}} \right)}^{\text{H}}}+\sum\limits_{l=1}^{L}{\frac{{{\mathbf{z}}_{l}}\mathbf{z}_{l}^{\text{H}}}{\hat{\sigma }_{l}^{1,({{n}_{\max }})}}} \right| \right.}
\end{equation}

\subsection{Rao}

The Rao test can be represented as \cite{LiuLiu2015d}
\begin{equation}\label{10}
	\begin{aligned}
		{{t}_{\text{Rao}}}=&\left. \frac{\partial \ln {{f}_{1}}(\mathbf{Z},{{\mathbf{Z}}_{L}})}{\partial {{\mathbf{\Theta }}_{r}}} \right|_{\mathbf{\Theta =}{{{\mathbf{\hat{\Theta }}}}_{\text{0}}}}^{\text{T}}{{\left[ {{\mathbf{F}}^{-1}}\left( {{{\mathbf{\hat{\Theta }}}}_{0}} \right) \right]}_{{{\mathbf{\Theta }}_{r}},{{\mathbf{\Theta }}_{r}}}}\\
		&\times{{\left. \frac{\partial \ln {{f}_{1}}(\mathbf{Z},{{\mathbf{Z}}_{L}})}{\partial \mathbf{\Theta }_{r}^{*}} \right|}_{\mathbf{\Theta }={{{\mathbf{\hat{\Theta }}}}_{0}}}}
	\end{aligned}
\end{equation}
For convenience, we denote $\mathbf{\Theta}$ as the complex-valued parameter vector, symbolized as
\begin{equation}
	\mathbf{\Theta }={{\left[ \mathbf{\Theta }_{r}^{\text{T}},\mathbf{\Theta }_{s}^{\text{T}} \right]}^{\text{T}}}
\end{equation}
with ${{\mathbf{\Theta }}_{r}}=\text{vec(}\mathbf{\Psi })$, ${{\mathbf{\Theta }}_{s}}={{\left[ {{\bm{\sigma }}^{\text{T}}},\text{ve}{{\text{c}}^{\text{T}}}(\mathbf{R}) \right]}^{\text{T}}}$, and they represent the related parameter and the uncorrelated parameter, respectively.
Specifically, ${{\mathbf{\hat{\Theta }}}_{0}}={{\left[ \mathbf{\hat{\Theta }}_{r\text{0}}^{\text{T}},\mathbf{\hat{\Theta }}_{s\text{0}}^{\text{T}} \right]}^{\text{T}}}$ represents the MLE of $\mathbf{\Theta}$ under hypothesis ${{\text{H}}_{0}}$ with ${{\mathbf{\hat{\Theta }}}_{r\text{0}}}={{\mathbf{0}}_{p\times K}}$ and ${{\mathbf{\hat{\Theta }}}_{s0}}={{\left[ \bm{\hat{\sigma }}_{0}^{\text{T}},\text{ve}{{\text{c}}^{\text{T}}}(\mathbf{\hat{R}}_{0}) \right]}^{\text{T}}}$.

Based on the partition matrix inversion theorem, we obtain
\begin{equation}\label{a11}
	\begin{aligned}
		{{[\mathbf{F}^{-1}(\mathbf{\Theta } )]}_{{{\mathbf{\Theta } }_{r}},{{\mathbf{\Theta } }_{r}}}}=&\left[ {{\mathbf{F}}_{{{\mathbf{\Theta } }_{r}},{{\mathbf{\Theta } }_{r}}}}(\mathbf{\Theta } )-{{\mathbf{F}}_{{{\mathbf{\Theta } }_{r}},{{\mathbf{\Theta } }_{s}}}}(\mathbf{\Theta } )\mathbf{F}_{{{\mathbf{\Theta } }_{s}},{{\mathbf{\Theta } }_{s}}}^{-1}(\mathbf{\Theta })\right.\\
		&\times\left.{{\mathbf{F}}_{{{\mathbf{\Theta } }_{s}},{{\mathbf{\Theta } }_{r}}}}(\mathbf{\Theta } ) \Big]^{-1}\right.
	\end{aligned}
\end{equation}%

It is recognized that the fisher information matrix (FIM) w.r.t. $\mathbf{\Theta}$ is expressible as
\begin{equation}\label{11}
	\mathbf{F}\left( \mathbf{\Theta } \right)=\text{E}\left[ \frac{\partial \ln f(\mathbf{Z};\mathbf{\Theta })}{\partial {{\mathbf{\Theta }}^{*}}}\frac{\partial \ln f(\mathbf{Z};\mathbf{\Theta })}{\partial {{\mathbf{\Theta }}^{\text{T}}}} \right]
\end{equation}

Calculating the partial derivative of the natural logarithm of \eqref{x3} w.r.t. $\mathbf{\Psi}$ and $\mathbf{\Psi}^{*}$, we can obtain
\begin{equation}\label{12}
	\frac{\partial \ln {{f}_{1}}\left( \mathbf{Z},{{\mathbf{Z}}_{L}} \right)}{\partial {{\mathbf{\Theta }}_{r}}}=\frac{\partial \ln {{f}_{1}}\left( \mathbf{Z},{{\mathbf{Z}}_{L}} \right)}{\partial \text{vec(}\mathbf{\Psi })}=\text{vec}\left[ {{\left( \mathbf{Z}_{1}^{\text{H}}{{\mathbf{R}}^{\mathbf{-1}}}\mathbf{H} \right)}^{\text{T}}} \right]
\end{equation}
and
\begin{equation}\label{13}
	\frac{\partial \ln {{f}_{1}}\left( \mathbf{Z},{{\mathbf{Z}}_{L}} \right)}{\partial \mathbf{\Theta }_{r}^{*}}=\frac{\partial \ln {{f}_{1}}\left( \mathbf{Z},{{\mathbf{Z}}_{L}} \right)}{\partial \text{vec(}\mathbf{\Psi }{{)}^{*}}}=\text{vec}\left( {{\mathbf{H}}^{\text{H}}}{{\mathbf{R}}^{-1}} \mathbf{Z}_{1}   \right)
\end{equation}
respectively. Substituting \eqref{12} and \eqref{13} into \eqref{11} results in 
\begin{equation}\label{14}
	\begin{aligned}
		{{\mathbf{F}}_{{{\mathbf{\Theta }}_{r}}\mathbf{,}{{\mathbf{\Theta }}_{r}}}}(\mathbf{\Theta }\text{)}=&\text{E}\left[ \frac{\partial \ln f(\mathbf{Z},{{\mathbf{Z}}_{L}})}{\partial \mathbf{\Theta }_{r}^{*}}\frac{\partial \ln f(\mathbf{Z},{{\mathbf{Z}}_{L}})}{\partial \mathbf{\Theta }_{r}^{\text{T}}} \right] \\ 
		=&\left( {{\mathbf{F}}_{K}}\otimes \left( {{\mathbf{H}}^{\text{H}}}{{\mathbf{R}}^{-1}} \right) \right)\text{E}\left[ \text{vec}\left( {{\mathbf{Z}}_{1}} \right)\text{ve}{{\text{c}}^{\text{H}}}\left( {{\mathbf{Z}}_{1}} \right) \right]\\
		&\times\left( {{\mathbf{F}}_{K}}\otimes \left( {{\mathbf{R}}^{-1}}\mathbf{H} \right) \right) \\ 
		=&{{\mathbf{F}}_{K}}\otimes {{\mathbf{H}}^{\text{H}}}{{\mathbf{R}}^{-1}}\mathbf{H} \\ 
	\end{aligned}
\end{equation}
and ${{\mathbf{F}}_{{{\mathbf{\Theta }}_{r}}\mathbf{,}{{\mathbf{\Theta }}_{s}}}}(\mathbf{\Theta }\text{)}$ can be written as
\begin{equation}\label{15}
	{{\mathbf{F}}_{{{\mathbf{\Theta }}_{r}}\mathbf{,}{{\mathbf{\Theta }}_{s}}}}(\mathbf{\Theta }\text{)=}\left[ {{\mathbf{F}}_{\text{vec(}\mathbf{\Psi }),\bm{\sigma }}}(\mathbf{\Theta }\text{)  }{{\mathbf{F}}_{\text{vec(}\mathbf{\Psi }),\text{vec}(\mathbf{R})}}(\mathbf{\Theta }\text{)} \right]
\end{equation}
Taking the derivative of \eqref{13} w.r.t. ${{\bm{\sigma }}^{\text{T}}}$ and $\text{ve}{{\text{c}}^{\text{T}}}(\mathbf{R})$, we obtain
\begin{equation}\label{16}
	{{\mathbf{F}}_{\text{vec(}\mathbf{\Psi }),\bm{\sigma }}}(\mathbf{\Theta })=-\text{E}\left[ \frac{{{\partial }^{2}}\ln {{f}_{1}}(\mathbf{Z},{{\mathbf{Z}}_{L}})}{\partial \text{vec(}\mathbf{\Psi }{{)}^{*}}\partial {{\bm{\sigma }}^{\text{T}}}} \right]
	={{\mathbf{0}}_{Kp\times L}}
\end{equation}
and
\begin{equation}\label{17}
	{{\mathbf{F}}_{\text{vec}(\mathbf{\Psi }),\text{vec}(\mathbf{R})}}(\mathbf{\Theta })=-\text{E}\left[ \frac{{{\partial }^{2}}\ln {{f}_{1}}(\mathbf{Z},{{\mathbf{Z}}_{L}})}{\partial \text{vec}{{(\mathbf{\Psi })}^{*}}\partial \text{ve}{{\text{c}}^{\text{T}}}(\mathbf{R})} \right]\\
	={{\mathbf{0}}_{Kp\times {{N}^{2}}}}
\end{equation}
respectively. Taking \eqref{14} and \eqref{15} into \eqref{a11} leads to
\begin{equation}\label{18}
	{{\left[ {{\mathbf{F}}^{-1}}(\mathbf{\Theta }\text{)} \right]}_{{{\mathbf{\Theta }}_{r}}{{\mathbf{\Theta }}_{r}}}}={{\mathbf{F}}_{K}}\otimes {{\left( {{\mathbf{H}}^{\text{H}}}{{\mathbf{R}}^{-1}}\mathbf{H} \right)}^{-1}}
\end{equation}
Substituting \eqref{12}, \eqref{13} and \eqref{18} into \eqref{10} yields AIE-Rao for the given $\mathbf{R}$ as
\begin{equation}\label{19}
	\begin{aligned}
		{{t}_{\text{Ra}{{\text{o}}_{\mathbf{R}}}}} =&\text{ve}{{\text{c}}^{\text{T}}}\left[ {{\left( \mathbf{Z}_{1}^{\text{H}}{{\mathbf{R}}^{-1}}\mathbf{H} \right)}^{\text{T}}} \right]\left[ {{\mathbf{F}}_{K}}\otimes {{\left( {{\mathbf{H}}^{\text{H}}}{{\mathbf{R}}^{-1}}\mathbf{H} \right)}^{-1}} \right]\\
		&\times\text{vec}\left( {{\mathbf{H}}^{\text{H}}}{{\mathbf{R}}^{-1}}{{\mathbf{Z}}_{1}} \right) \\ 
		=&{{\left[ \left( {{\mathbf{F}}_{K}}\otimes {{\mathbf{H}}^{\text{T}}}{{\mathbf{R}}^{-\text{T}}} \right)\text{vec}\left( \mathbf{Z}_{1}^{*} \right) \right]}^{\text{T}}}\left[ {{\mathbf{F}}_{K}}\otimes {{\left( {{\mathbf{H}}^{\text{H}}}{{\mathbf{R}}^{-1}}\mathbf{H} \right)}^{-1}} \right]\\
		&\times\left( {{\mathbf{F}}_{K}}\otimes {{\mathbf{H}}^{\text{H}}}{{\mathbf{R}}^{-1}} \right)\text{vec}\left( {{\mathbf{Z}}_{1}} \right) \\ 
		=&\text{ve}{{\text{c}}^{\text{H}}}\left( {{\mathbf{Z}}_{1}} \right)\left( {{\mathbf{F}}_{K}}\otimes {{\mathbf{R}}^{-1}}\mathbf{H} \right)\left[ {{\mathbf{F}}_{K}}\otimes {{\left( {{\mathbf{H}}^{\text{H}}}{{\mathbf{R}}^{-1}}\mathbf{H} \right)}^{-1}} \right]\\
		&\times\left( {{\mathbf{F}}_{K}}\otimes {{\mathbf{H}}^{\text{H}}}{{\mathbf{R}}^{-1}} \right)\text{vec}\left( {{\mathbf{Z}}_{1}} \right) \\ 
		=&\text{ve}{{\text{c}}^{\text{H}}}\left( {{\mathbf{Z}}_{1}} \right)\left\{ {{\mathbf{F}}_{K}}\otimes \left[ {{\mathbf{R}}^{-1}}\mathbf{H}{{\left( {{\mathbf{H}}^{\text{H}}}{{\mathbf{R}}^{-1}}\mathbf{H} \right)}^{-1}}{{\mathbf{H}}^{\text{H}}}{{\mathbf{R}}^{-1}} \right] \right\}\\ 
		&\times\text{vec}\left( {{\mathbf{Z}}_{1}} \right)\\
		=&\text{ve}{{\text{c}}^{\text{H}}}\left( {{\mathbf{Z}}_{1}} \right)\text{vec}\left[ {{\mathbf{R}}^{-1}}\mathbf{H}{{\left( {{\mathbf{H}}^{\text{H}}}{{\mathbf{R}}^{-1}}\mathbf{H} \right)}^{-1}}{{\mathbf{H}}^{\text{H}}}{{\mathbf{R}}^{-1}}{{\mathbf{Z}}_{1}} \right] \\ 
		=&\text{tr}\left[ \mathbf{Z}_{1}^{\text{H}}{{\mathbf{R}}^{-1}}\mathbf{H}{{\left( {{\mathbf{H}}^{\text{H}}}{{\mathbf{R}}^{-1}}\mathbf{H} \right)}^{-1}}{{\mathbf{H}}^{\text{H}}}{{\mathbf{R}}^{-1}}{{\mathbf{Z}}_{1}} \right] \\ 
		=&\text{tr}\left[ {{\mathbf{Z}}^{\text{H}}}{{\mathbf{R}}^{-1}}\mathbf{H}{{\left( {{\mathbf{H}}^{\text{H}}}{{\mathbf{R}}^{-1}}\mathbf{H} \right)}^{-1}}{{\mathbf{H}}^{\text{H}}}{{\mathbf{R}}^{-1}}\mathbf{Z} \right] \\ 
	\end{aligned}
\end{equation}
where the last equality makes use of $\mathbf{\Psi }={{\mathbf{0}}_{p\times K}}$.

Under hypothesis ${{\text{H}}_{0}}$, substituting the ultimate estimation of $\bm{\sigma}$ into \eqref{x6} yields the MLE of $\mathbf{R}$ as
\begin{equation}\label{20}
	{{\mathbf{\hat{R}}}_{0}}=\frac{1}{L+K}\left( \mathbf{Z}{{\mathbf{Z}}^{\text{H}}}+\sum\limits_{l=1}^{L}{\frac{{{\mathbf{z}}_{l}}\mathbf{z}_{l}^{\text{H}}}{\hat{\sigma} _{l}^{0,\left( {{n}_{\max }} \right)}}} \right)
\end{equation}
For the sake of brevity, we let ${{\mathbf{\Xi }}_{0}}=\mathbf{Z}{{\mathbf{Z}}^{\text{H}}}+\sum\limits_{l=1}^{L}{\frac{{{\mathbf{z}}_{l}}\mathbf{z}_{l}^{\text{H}}}{\hat{\sigma} _{l}^{0,\left( {{n}_{\max }} \right)}}}$. Then taking \eqref{20} into \eqref{19} and setting aside the constant terms, we can derive the AIE-Rao as
\begin{equation}\label{21}
	{{t}_{\text{AIE-Rao}}}=\text{tr}\left[ {{\mathbf{Z}}^{\text{H}}}\mathbf{\Xi }_{0}^{-1}\mathbf{H}{{\left( {{\mathbf{H}}^{\text{H}}}\mathbf{\Xi }_{0}^{-1}\mathbf{H} \right)}^{-1}}{{\mathbf{H}}^{\text{H}}}\mathbf{\Xi }_{0}^{-1}\mathbf{Z} \right]
\end{equation}

\subsection{Wald}
The Wald test can be represented as \cite{Liu20SCIS}
\begin{equation}\label{22}
	{{t}_{\text{Wald}}}={{\left( {{{\mathbf{\hat{\Theta }}}}_{r1}}-{{\mathbf{\Theta }}_{r0}} \right)}^{\text{H}}}{{\left\{ {{\left[ {{\mathbf{F}}^{-1}}\left( {{{\mathbf{\hat{\Theta }}}}_{1}} \right) \right]}_{{{\mathbf{\Theta }}_{r}},{{\mathbf{\Theta }}_{r}}}} \right\}}^{-1}}\left( {{{\mathbf{\hat{\Theta }}}}_{r1}}-{{\mathbf{\Theta }}_{r0}} \right)
\end{equation}
where ${{\mathbf{\hat{\Theta }}}_{1}}={{\left[ \mathbf{\hat{\Theta }}_{r\text{1}}^{\text{T}},\mathbf{\hat{\Theta }}_{s\text{1}}^{\text{T}} \right]}^{\text{T}}}$ denotes the MLE of $\mathbf{\Theta}$ under hypothesis ${{\text{H}}_{1}}$ with ${{\mathbf{\hat{\Theta }}}_{r\text{1}}}=\text{vec}\left( \mathbf{\hat{\Psi }} \right)$ and ${{\mathbf{\hat{\Theta }}}_{s1}}={{\left[ \bm{\hat{\sigma }}_{1}^{\text{T}},\text{ve}{{\text{c}}^{\text{T}}}(\mathbf{\hat{R}}_{1}) \right]}^{\text{T}}}$ and ${{\mathbf{\Theta }}_{r0}}$ represents the value of $\mathbf{{\Theta}}_{r}$ under hypothesis ${{\text{H}}_{0}}$. Letting the derivative of \eqref{x3} w.r.t. $\mathbf{\Psi }$ be zero, we can derive the MLE of $\mathbf{\Psi }$ for the given $\mathbf{R}$ as
\begin{equation}\label{23}
	\mathbf{\hat{\Psi }}={{\left( {{\mathbf{H}}^{\text{H}}}{{\mathbf{R}}^{-1}}\mathbf{H} \right)}^{-1}}{{\mathbf{H}}^{\text{H}}}{{\mathbf{R}}^{-1}}\mathbf{Z}
\end{equation}
taking \eqref{18} and \eqref{23} into \eqref{22} results in the AIE-Wald for the given $\mathbf{R}$ as
\begin{equation}\label{24}
	\begin{aligned}
		{{t}_{\text{Wald}_{\mathbf{R}}}}= &\text{vec}^{\text{H}}\left[ {{\left( {{\mathbf{H}}^{\text{H}}}{{\mathbf{R}}^{-1}}\mathbf{H} \right)}^{-1}}{{\mathbf{H}}^{\text{H}}}{{\mathbf{R}}^{-1}}\mathbf{Z} \right]\left[ {{\mathbf{F}}_{K}}\otimes {{\mathbf{H}}^{\text{H}}}{{\mathbf{R}}^{-1}}\mathbf{H} \right]\\
		&\times\text{vec}\left[ {{\left( {{\mathbf{H}}^{\text{H}}}{{\mathbf{R}}^{-1}}\mathbf{H} \right)}^{-1}}{{\mathbf{H}}^{\text{H}}}{{\mathbf{R}}^{-1}}\mathbf{Z} \right] \\ 
		= &\text{ve}{{\text{c}}^{\text{H}}}\left[ {{\left( {{\mathbf{H}}^{\text{H}}}{{\mathbf{R}}^{-1}}\mathbf{H} \right)}^{-1}}{{\mathbf{H}}^{\text{H}}}{{\mathbf{R}}^{-1}}\mathbf{Z} \right]\text{vec}\left( {{\mathbf{H}}^{\text{H}}}{{\mathbf{R}}^{-1}}\mathbf{Z} \right) \\ 
		=&\text{tr}\left[ {{\mathbf{Z}}^{\text{H}}}{{\mathbf{R}}^{-1}}\mathbf{H}{{\left( {{\mathbf{H}}^{\text{H}}}{{\mathbf{R}}^{-1}}\mathbf{H} \right)}^{-1}}{{\mathbf{H}}^{\text{H}}}{{\mathbf{R}}^{-1}}\mathbf{Z} \right] \\ 
	\end{aligned}
\end{equation}
Similarly, under hypothesis ${{\text{H}}_{1}}$, taking the ultimate estimation of $\bm{\sigma}$ and $\mathbf{\Psi}$ into \eqref{x5} yields the MLE of $\mathbf{R}$ as
\begin{equation}\label{25}
	{{\mathbf{\hat{R}}}_{1}}=\frac{\left[ \left( \mathbf{Z}-\mathbf{H}{{{\mathbf{\hat{\Psi }}}}^{\left( {{n}_{\max }} \right)}} \right){{\left( \mathbf{Z}-\mathbf{H}{{{\mathbf{\hat{\Psi }}}}^{\left( {{n}_{\max }} \right)}} \right)}^{\text{H}}}+\sum\limits_{l=1}^{L}{\frac{{{\mathbf{z}}_{l}}\mathbf{z}_{l}^{\text{H}}}{\hat{\sigma} _{l}^{1,\left( {{n}_{\max }} \right)}}} \right]}{L+K}
\end{equation}
Defining ${{\mathbf{\Xi }}_{1}}=\left( \mathbf{Z}-\mathbf{H}{{{\mathbf{\hat{\Psi }}}}^{\left( {{n}_{\max }} \right)}} \right){{\left( \mathbf{Z}-\mathbf{H}{{{\mathbf{\hat{\Psi }}}}^{\left( {{n}_{\max }} \right)}} \right)}^{\text{H}}}+\sum\limits_{l=1}^{L}{\frac{{{\mathbf{z}}_{l}}\mathbf{z}_{l}^{\text{H}}}{\hat{\sigma} _{l}^{1,\left( {{n}_{\max }} \right)}}}$, taking \eqref{25} into \eqref{24}, and disregarding the constant terms yields the AIE-Wald as
\begin{equation}
	{{t}_{\text{AIE-Wald}}}=\text{tr}\left[ {{\mathbf{Z}}^{\text{H}}}\mathbf{\Xi }_{1}^{-1}\mathbf{H}{{\left( {{\mathbf{H}}^{\text{H}}}\mathbf{\Xi }_{1}^{-1}\mathbf{H} \right)}^{-1}}{{\mathbf{H}}^{\text{H}}}\mathbf{\Xi }_{1}^{-1}\mathbf{Z} \right]
\end{equation}

\section{Performance Evaluation}
In this section, we evaluate the performance of the proposed detectors by the use of both simulated and real data. Set the predefined value of ${\sigma}_{l}$ to be 
$\hat{\sigma }_{l}^{(0)}=\mathbf{z}_{l}^{\text{H}}{{\left( \mathbf{Z}{{\mathbf{Z}}^{\text{H}}} \right)}^{-}}{{\mathbf{z}}_{l}},\text{  }l=1,2, \cdots,L$  \cite{Coluccia2022}. According to the deterministic multiple dominant scattering (MDS) model \cite{AlfanoRSN04}, we consider the target energy uniform scattering model in experiment. Specifically, this model refers to the case that the scattered echo energy of radar targets is uniformly distributed in each range bin.

As the competitor to our proposed detectors, we consider the ANMF \cite{ConteLops95}, which performs well in non-Gaussian environments, combined with the recursive estimation (RE) for the unknown covariance matrix. The RE is computed through an iterative procedure as \cite{ConteDeMaio02b}
\begin{equation}\label{x37}
	\mathbf{\Omega}(r+1)=\frac{N}{L}\sum\limits_{l=1}^{L}{\frac{{{\mathbf{z}}_{l}}\mathbf{z}_{l}^{\text{H}}}{\mathbf{z}_{l}^{\text{H}}\mathbf{\hat{R}}_{\text{RE}}^{-1}(r){{\mathbf{z}}_{l}}}}
\end{equation}
and
\begin{equation}\label{x38}
	{{\mathbf{\hat{R}}}_{\text{RE}}}(r+1)=N\frac{\mathbf{\Omega}\left( r+1 \right)}{\text{tr}\left[ \mathbf{\Omega}\left( r+1 \right) \right]}
\end{equation}
where the number of iteration for the above RE is set to be 4 to ensure good estimation accuracy\cite{ConteDeMaio02b}. Based on \eqref{x37} and \eqref{x38}, the expression for ANMF with the RE is
\begin{equation}
	{{t}_{\text{ANMF-RE}}}=\frac{\text{tr}\left[ {{\mathbf{Z}}^{\text{H}}}\mathbf{\hat{R}}_{\text{RE}}^{-1}\mathbf{H}{{\left( {{\mathbf{H}}^{\text{H}}}\mathbf{\hat{R}}_{\text{RE}}^{-1}\mathbf{H} \right)}^{-1}}{{\mathbf{H}}^{\text{H}}}\mathbf{\hat{R}}_{\text{RE}}^{-1}\mathbf{Z} \right]}{\text{tr}\left( {{\mathbf{Z}}^{\text{H}}}\mathbf{\hat{R}}_{\text{RE}}^{-1}\mathbf{Z} \right)}
\end{equation}

We set the PFA as ${{10}^{-3}}$ and conduct ${{10}^{4}}$ independent repeated trials to compute probability of detection (PD). To obtain an appropriate detection threshold according to the predefined PFA, we run $100/\text{PFA}$ Monte Carlo experiments.

\begin{figure}[!t]
\centerline{\includegraphics[width=\columnwidth]{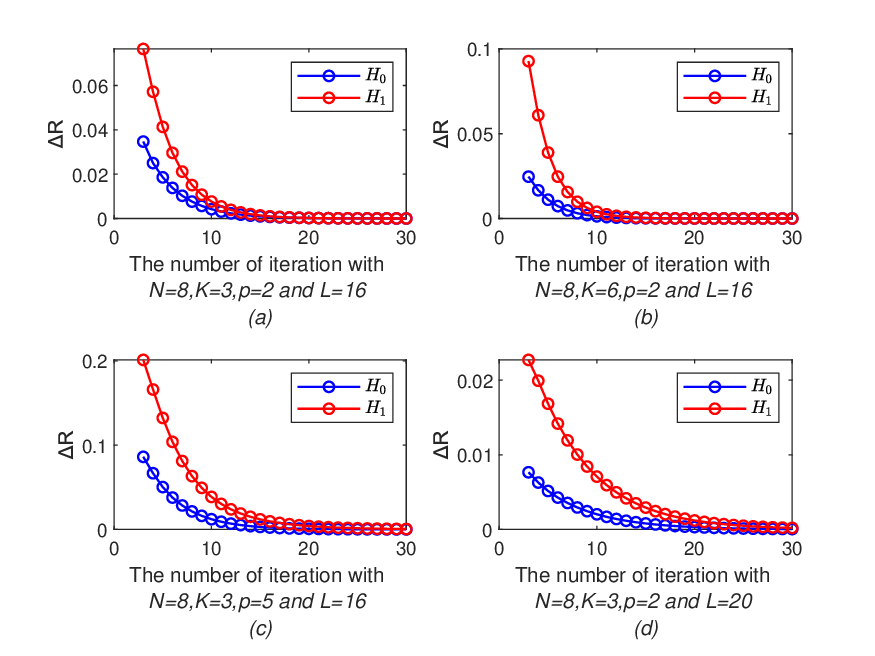}}
\caption{The relative change in clutter covariance matrix between consecutive iterations.}
\label{fig_error}
\end{figure}

In order to set a reasonable maximum number of iterations, we define the variation of covariance matrix $\mathbf{R}$ occurring between two consecutive iterations as
\begin{equation}
	\Delta R=\frac{\left| \left. {{\left\| {{{\mathbf{\hat{R}}}}^{(n+1)}} \right\|}_{\text{F}}}-{{\left\| {{{\mathbf{\hat{R}}}}^{(n)}} \right\|}_{\text{F}}} \right| \right.}{{{\left\| {{{\mathbf{\hat{R}}}}^{(n)}} \right\|}_{\text{F}}}}
\end{equation}

\begin{figure}[!t]
	\centerline{\includegraphics[width=\columnwidth]{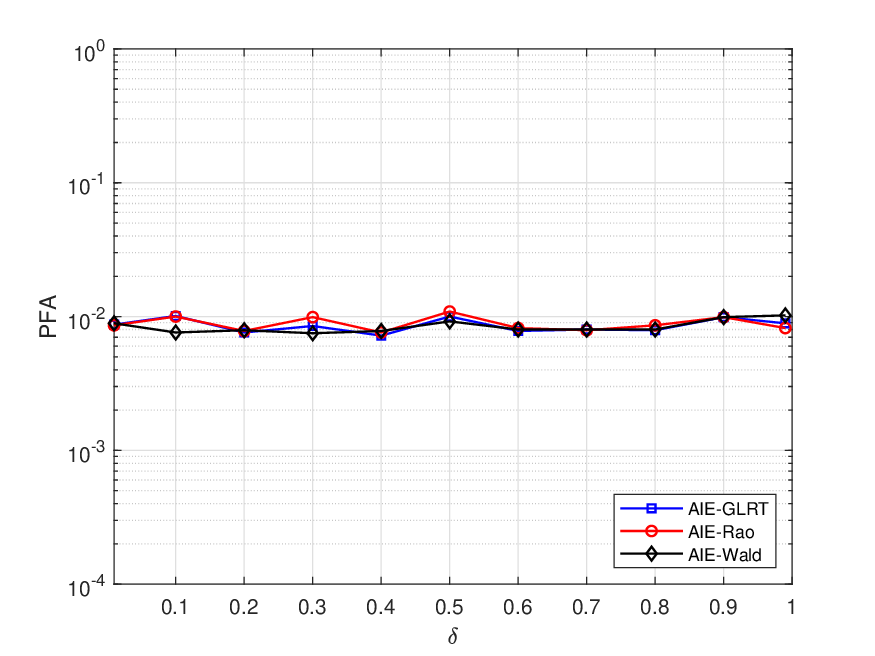}}
	\caption{PFAs versus $\delta$ of clutter covariance matrix of proposed detectors with simulated data.}
	\label{fig_cfar}
\end{figure}

\begin{figure}[!htp]
	\centering
	\subfigure[$\text{Simulated data}$]{\includegraphics[width=\columnwidth]{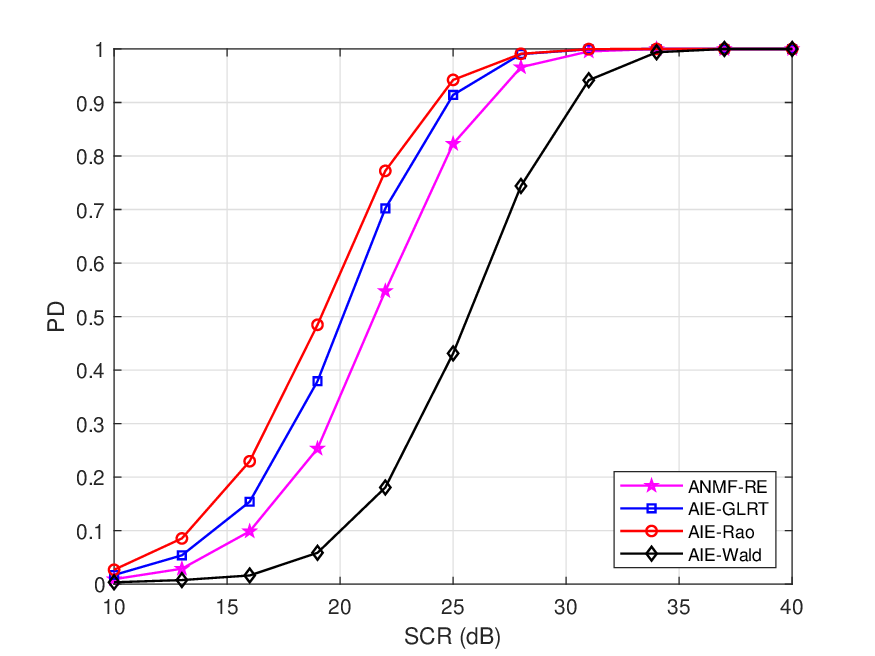}}
	\subfigure[$\text{IPIX radar data}$]{\includegraphics[width=\columnwidth]{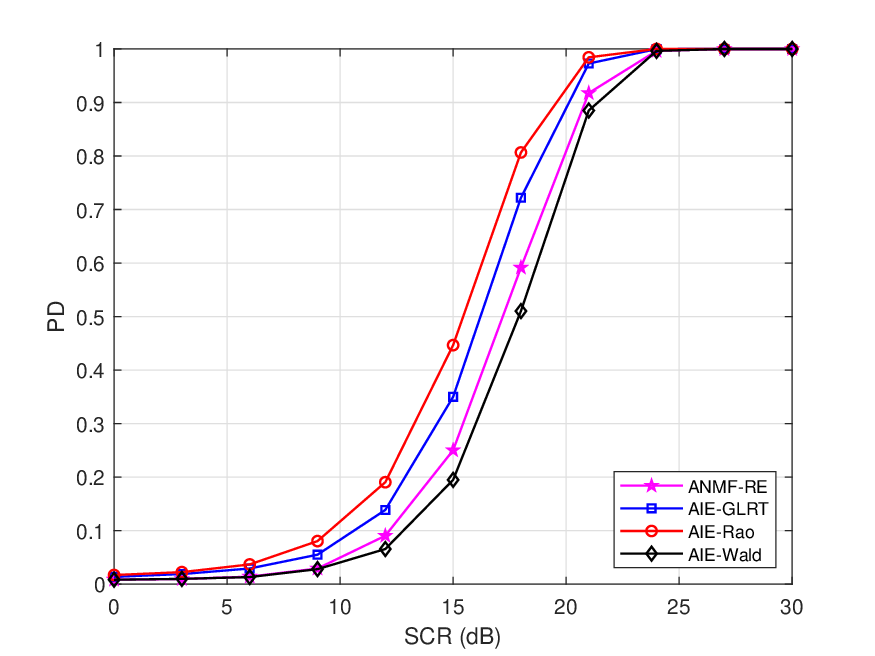}}
	\caption{PD varies with SCR in simulated and real data for $N=8$, $K=3$, $p=2$ and $L=16$.}
	\label{Fig1}
\end{figure}

\begin{figure}
	\centering
	\subfigure[$\text{Simulated data}$]{\includegraphics[width=\columnwidth]{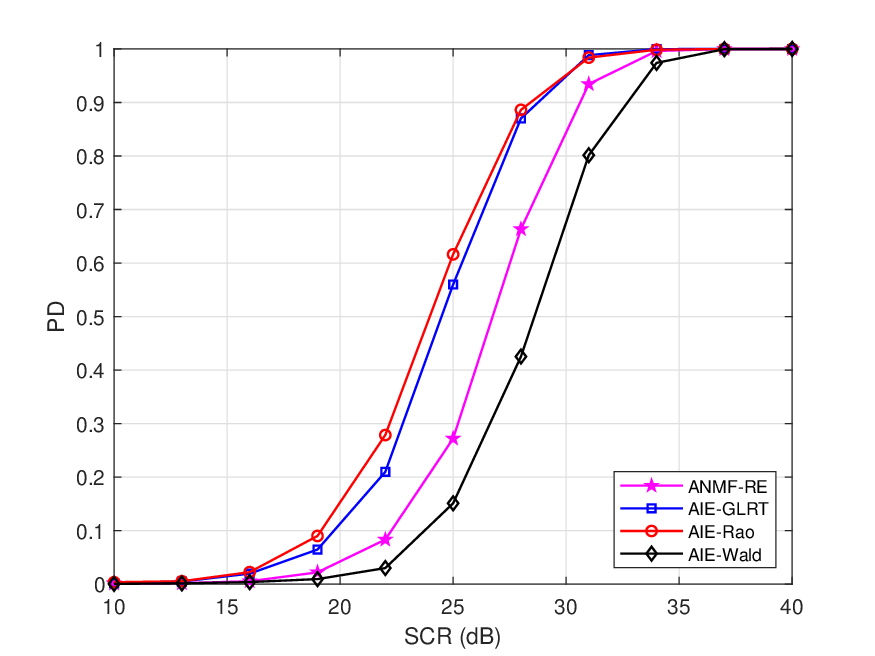}}
	\subfigure[$\text{IPIX radar data}$]{\includegraphics[width=\columnwidth]{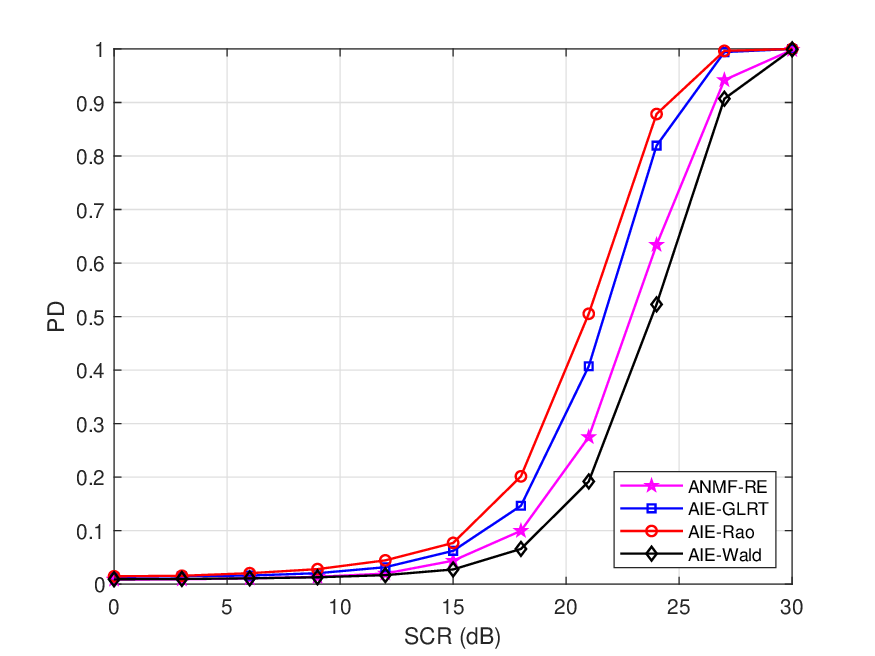}}
	\caption{PD varies with SCR in simulated and real data for $N=8$, $K=6$, $p=2$ and $L=16$.}
	\label{Fig2}
\end{figure}

\begin{figure}
	\centering
	\subfigure[$\text{Simulated data}$]{\includegraphics[width=\columnwidth]{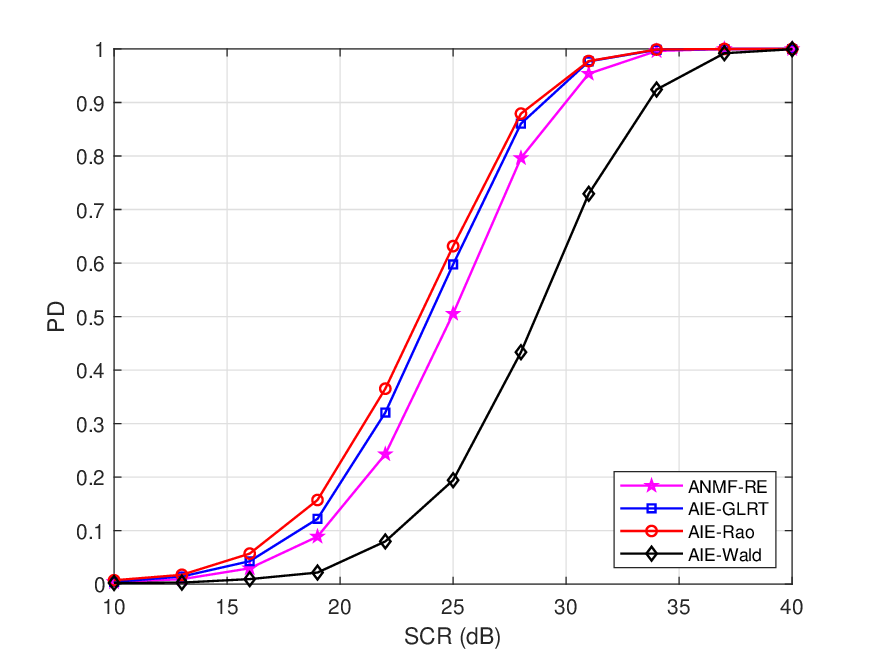}}
	\subfigure[$\text{IPIX radar data}$]{\includegraphics[width=\columnwidth]{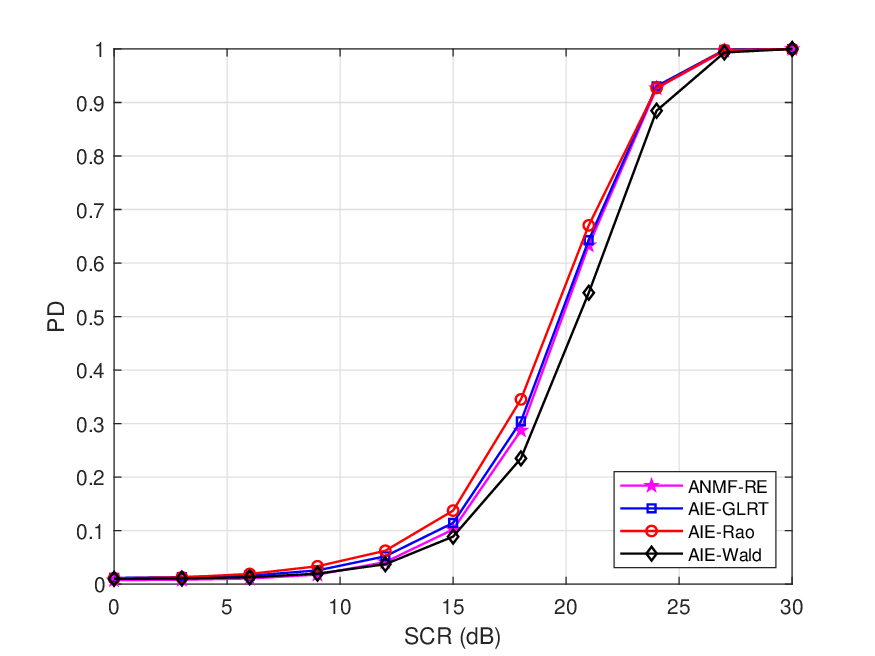}}
	\caption{PD varies with SCR in simulated and real data for $N=8$, $K=3$, $p=5$ and $L=16$.}
	\label{Fig3}
\end{figure}

\begin{figure}
	\centering
	\subfigure[$\text{Simulated data}$]{\includegraphics[width=\columnwidth]{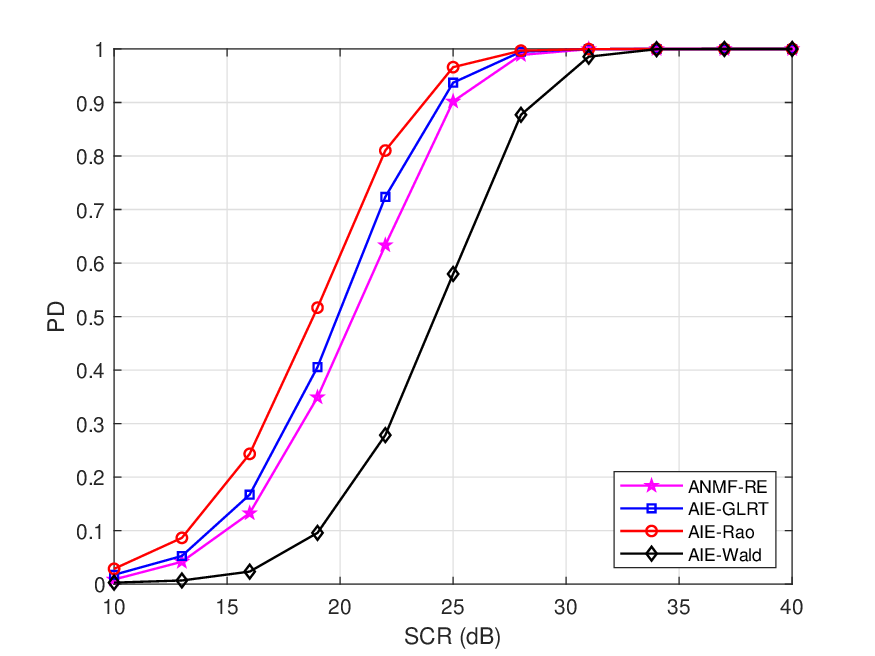}}
	\subfigure[$\text{IPIX radar data}$]{\includegraphics[width=\columnwidth]{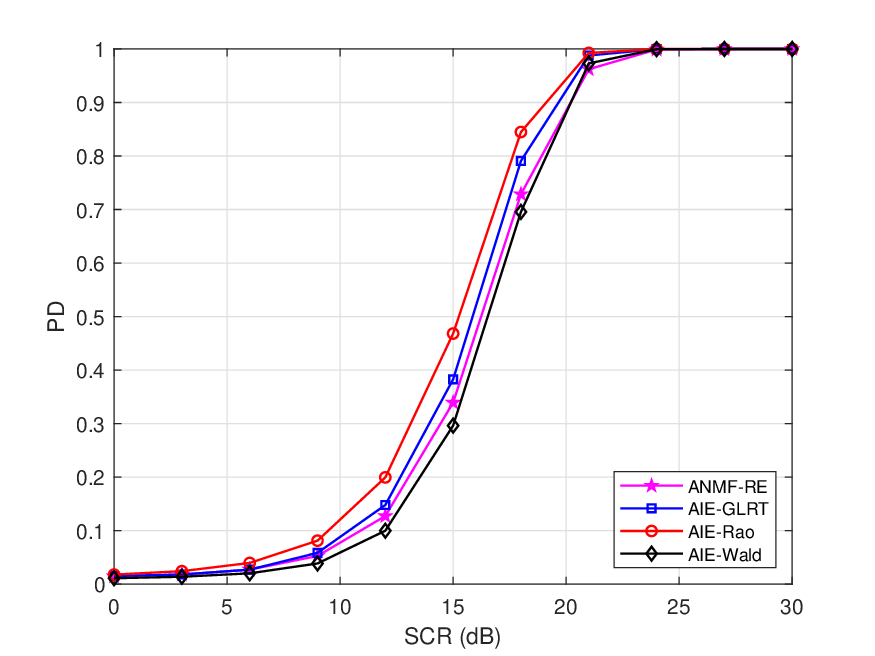}}
	\caption{PD varies with SCR in simulated and real data for $N=8$, $K=3$, $p=2$ and $L=20$.}
	\label{Fig4}
\end{figure}

Fig. \ref{fig_error} plots the relative change in covariance matrix $\mathbf{R}$ during successive iterations under different parameter settings. In particular, Fig. \ref{fig_error} (a), (b), (c) and (d) correspond to the parameter settings in Fig. \ref{Fig1}, Fig. \ref{Fig2}, Fig. \ref{Fig3} and Fig. \ref{Fig4}, respectively. Obviously, with the number of iteration increases, the relative variation in the estimate of $\mathbf{R}$ decreases. Specifically, upon reaching $15$ iterations, the relative variation gradually tends to stable. Taking into account the complexity of computation and the performance of proposed iterative approach, we set the number of alternating iterative estimates to be 15.

In simulation experiments, we generate K-distributed clutter, with PDF of texture component being \cite{JiangWu23}
\begin{equation}
	p(\tau )=\frac{{{\tau }^{v-1}}}{\Gamma (v)}{{\left( \frac{v}{u} \right)}^{v}}\exp \left( -\frac{v}{u}\tau  \right)
\end{equation}
where $v$ represents shape parameter. A lower $v$ indicates a more pronounced non-Gaussian characteristic, and $u$ represents scale parameter, used to quantify the mean power of clutter. We set $v$ and $u$ both to be 1 in the simulated experiment.

In the experiment, we set signal-to-clutter ratio (SCR) as
\begin{equation}\label{43}
	\text{SC}{{\text{R}}}=\text{tr}\left( {{\mathbf{\Psi }}^{\text{H}}}{{\mathbf{H}}^{\text{H}}}{{\mathbf{\Sigma}}^{-1}}\mathbf{H\Psi } \right)
\end{equation}
The value of the $(a,b)$-th element in the speckle covariance matrix is ${{\left[ \mathbf{\Sigma} \right]}_{a,b}}={{\delta }^{\left| \left. a-b \right| \right.}}\text{,  }1\le a,b\le N$, where $\delta $ represents one-lag correlation coefficient and set $\delta$ to be $0.95$.

The following introduces the IPIX dataset used in the real data experiments of this paper. The IPIX radar sea clutter measurement experiment was conducted by McMaster University in Canada, and the dataset from the measurements was managed and maintained by a team led by Professor Haykin, representing typical shore-based platform data with a small grazing angle sea clutter. The experiment is widely known in the field of radar sea target detection technology, and the experimental data has been widely applied. For example, the IPIX database has become an important tool for evaluating the performance of radar detectors.

 As we all known, the IPIX dataset consists of two major collections in 1993 and 1998. The former is the measurement data from Dartmouth, a city in the southern part of Nova Scotia, Canada, in 1993. During the test, the radar overlooked the Atlantic Ocean from a cliff, with the latitude/longitude at 44°36.72N/63°25.41W, and the setup was about 30 meters above mean sea level. The latter is the measurement data from Grimsby, on the shores of Lake Ontario, Canada, in 1998. During the test, the radar overlooked Lake Ontario from the shore, with the latitude/longitude at 43°12'41.0"N/79°35'54.6"W, and the platform was at a height of 20 meters. The radar operated in dwell mode, the frequency is approximately 9.39 GHz with a peak power of 8 kW and a bandwidth of 50 MHz. The pulse repetition frequency is 1 kHz, the beam width is approximately 1.1 degrees, and four polarization types are acquired: HH, HV, VH, and VV polarizations. 

In real data experiment, we utilize the IPIX 1998 dataset file 86, named \text{"86dataHH"}, which contains 60000 pulses and 34 range bins, to evaluate the performance of proposed detectors. Specifically, we select the $\left(12,11+K \right)$ range bins as CUT, and choose the $\left(12-L/2,11 \right)$ and $\left(12+K,11+K+L/2 \right)$ range bins close to CUT as training samples. The matrix $\mathbf{\Sigma}$ in \eqref{43} is estimated by all pulses in the \text{10-th} range bin near the CUT. Because of the limited quantity of pulses in the real data, we set PFA as ${10}^{-2}$ and all other parameters are consistent with the simulated experiment.

Fig. \ref{fig_cfar} plots the variation of PFAs with $\delta$ of clutter covariance matrix. The results demonstrate that AIE-GLRT, AIE-Rao and AIE-Wald are all CFAR to the structure of clutter covariance matrix. A theoretical derivation of the CFAR property for point target is given in \cite{Coluccia2022}.

Fig. \ref{Fig1} plots the variation of PDs with SCR and analyzes the detection performance with simulated and real data. The graphs indicate that the results of the simulation experiment basically align with those of the real data experiment. Moreover, the AIE-GLRT and AIE-Rao both have higher PDs than the ANMF, and AIE-Rao exhibits the best performance.

Fig. \ref{Fig2} displays the variation of PDs with SCR and compares the detection performance using simulated and real data. We have increased the extended dimension of the target. It can be found that AIE-Rao has the highest PD. Moreover, the PDs of all detectors diminish as the extended dimension $K$ increases. A reasonable explanation is that as the target extended dimension increases, the size of the unknown parameter space expands, resulting in more dispersed target energy distribution, which causes the degradation in detection performance.

Fig. \ref{Fig3} plots the variation of PDs with SCR and evaluates the detection performance with simulated and real data. We have increased the dimension of the signal subspace. Obviously, AIE-Rao has the best performance in almost all cases. As the subspace dimension becomes larger, correspondingly, the dimension of the signal matrix also increases, leading to a greater number of parameters that need to be estimated and the uncertainty of the signal coordinates increases, which may cause some loss in detection performance.

Fig. \ref{Fig4} gives the variation of PDs with SCR and compares the detection performance through simulated and real data. We have increased the number of training samples. The proposed AIE-Rao has been noted to exhibit a higher PD compared with other detectors. Moreover, with the number of training samples increases, the PDs of all detectors enhance, as a consequence of the more accurate estimate of clutter covariance matrix.

\section{Conclusion}
This paper considered the adaptive detection of subspace-based distributed targets in power heterogeneous clutter. Based on GLRT, Rao and the Wald tests, we proposed AIE-GLRT, AIE-Rao and AIE-Wald through alternating iterative estimation. Simulation experiment and real data both demonstrated that AIE-Rao and AIE-GLRT exhibited higher PDs compared with ANMF, and the AIE-Rao showed the best performance. Our research further indicated that the performance of the proposed detectors improved through the reduction in the dimension of signal subspace, the decrease in the dimension of range expansion, and the increase of training samples. In addition, simulation experiment also revealed that the AIE-GLRT, AIE-Rao and AIE-Wald were all CFAR to the structure of clutter covariance matrix. This paper does not consider the detection problem in the presence of interference, which usually arises in practice, intentionally or unintentionally. Hence, considering the distributed target detection in power heterogeneous clutter plus subspace interference may be more practical and is a direction for future work.

\section*{APPENDIX}
The appendix provides the detailed derivations of the alternating iterative estimation of $\bm{\sigma}$ under hypotheses $\text{H}_{1}$ and $\text{H}_{0}$. Firstly, we define the following function
\begin{equation}\label{x13}
	\begin{aligned}
		f({{\sigma }_{h}})=& \sigma _{h}^{N}{{\left| \left.(\mathbf{Z}-\mathbf{H}{{{\bm{\hat{\Psi }}}}^{(n+1)}}){{(\mathbf{Z}-\mathbf{H}{{{\bm{\hat{\Psi }}}}^{(n+1)}})}^{\text{H}}}+\sum\limits_{l=1}^{h-1}{\frac{{{\mathbf{z}}_{l}}\mathbf{z}_{l}^{\text{H}}}{\hat{\sigma }_{l}^{1,(n+1)}}}\right.\right.}}\\
		&\left.\left.{+\sum\limits_{l=h+1}^{L}{\frac{{{\mathbf{z}}_{l}}\mathbf{z}_{l}^{\text{H}}}{\hat{\sigma }_{l}^{1,(n)}}+\frac{{{\mathbf{z}}_{h}}\mathbf{z}_{h}^{\text{H}}}{{{\sigma }_{h}}}}} \right| \right.^{L+K} \\ 
		=& \sigma _{h}^{N}{{\left| \left. \mathbf{\Lambda}_{h}^{1,(n+1)}+\frac{{{\mathbf{z}}_{h}}\mathbf{z}_{h}^{\text{H}}}{{{\sigma }_{h}}} \right| \right.}^{L+K}} \\ 
		=& \sigma _{h}^{N}{{\left| \left. \mathbf{\Lambda}_{h}^{1,(n+1)} \right| \right.}^{L+K}}{{\left( 1+\frac{\mathbf{z}_{h}^{\text{H}}{{\left[ \mathbf{\Lambda}_{h}^{1,(n+1)} \right]}^{-1}}{{\mathbf{z}}_{h}}}{{{\sigma }_{h}}} \right)}^{L+K}} \\ 
	\end{aligned}
\end{equation}
where
\begin{equation}\label{x14}
	\begin{aligned}
		\mathbf{\Lambda }_{h}^{1,\left( n+1 \right)}=&(\mathbf{Z}-\mathbf{H}{{\mathbf{\hat{\Psi }}}^{(n+1)}}){{(\mathbf{Z}-\mathbf{H}{{\mathbf{\hat{\Psi }}}^{(n+1)}})}^{\text{H}}}+\sum\limits_{l=1}^{h-1}{\frac{{{\mathbf{z}}_{l}}\mathbf{z}_{l}^{\text{H}}}{\hat{\sigma }_{l}^{1,(n+1)}}}\\
		&+\sum\limits_{l=h+1}^{L}{\frac{{{\mathbf{z}}_{l}}\mathbf{z}_{l}^{\text{H}}}{\hat{\sigma }_{l}^{1,(n)}}}
	\end{aligned}
\end{equation}
and we have used \eqref{x10}. It can be observed that the value of $f({{\sigma }_{h}})$ tends to infinity, as $\mathbf{\sigma}_{h}$ tends towards zero or infinity. Therefore, we can conclude that $f({{\sigma }_{h}})$ attains its minimum at a certain value within the range of 0 to $\infty$. Calculating the partial derivative of \eqref{x13} w.r.t. ${{\sigma }_{h}}$ results in
\begin{equation}\label{x15}
	\begin{aligned}
		& \frac{\partial f({{\sigma }_{h}})}{\partial {{\sigma }_{h}}}=N{{\sigma }_{h}}^{N-1}{{\left| \left. \mathbf{\Lambda }_{h}^{1,(n+1)} \right| \right.}^{L+K}}{{\left( 1+\frac{\mathbf{z}_{h}^{\text{H}}{{\left[ \mathbf{\Lambda }_{h}^{1,(n+1)} \right]}^{-1}}{{\mathbf{z}}_{h}}}{{{\sigma }_{h}}} \right)}^{L+K}} \\ 
		& \text{              }+\sigma _{h}^{N}{{\left| \left. \mathbf{\Lambda }_{h}^{1,(n+1)} \right| \right.}^{L+K}}\left( L+K \right){{\left( 1+\frac{\mathbf{z}_{h}^{\text{H}}{{\left[ \mathbf{\Lambda }_{h}^{1,(n+1)} \right]}^{-1}}{{\mathbf{z}}_{h}}}{{{\sigma }_{h}}} \right)}^{L+K-1}} \\ 
		& \text{            }\times \left( -\frac{1}{{{\sigma }_{h}}^{2}} \right)\left( \mathbf{z}_{h}^{\text{H}}{{\left[ \mathbf{\Lambda }_{h}^{1,(n+1)} \right]}^{-1}}{{\mathbf{z}}_{h}} \right) \\ 
	\end{aligned}
\end{equation}
By setting \eqref{x15} to be zero, we can obtain
\begin{equation}\label{x16}
	\hat{\sigma }_{h}^{1,(n+1)}=\frac{L+K-N}{N}\left\{ \mathbf{z}_{h}^{\text{H}}{{\left[ \mathbf{\Lambda }_{h}^{1,(n+1)} \right]}^{-1}}{{\mathbf{z}}_{h}} \right\}
\end{equation}
where $h=1,2,\cdots ,L$.

It is evident that $\mathbf{\Psi }={{\mathbf{0}}_{p\times K}}$ under hypothesis $\text{H}_{0}$. Similarly, We provide the value of alternating iterative estimation of $\bm{\sigma}$ under hypothesis $\text{H}_{0}$ as
\begin{equation}\label{x17}
	\hat{\sigma }_{h}^{0,(n+1)}=\frac{L+K-N}{N}\left( \mathbf{z}_{h}^{\text{H}}{{\left[ \mathbf{\Lambda}_{h}^{0,(n+1)} \right]}^{-1}}{{\mathbf{z}}_{h}} \right)
\end{equation}
where
\begin{equation}\label{x18}
	\mathbf{\Lambda}_{h}^{0,\left( n+1 \right)}=\mathbf{Z}{{\mathbf{Z}}^{\text{H}}}+\sum\limits_{l=1}^{h-1}{\frac{{{\mathbf{z}}_{l}}\mathbf{z}_{l}^{\text{H}}}{\hat{\sigma }_{l}^{0,(n+1)}}}+\sum\limits_{l=h+1}^{L}{\frac{{{\mathbf{z}}_{l}}\mathbf{z}_{l}^{\text{H}}}{\hat{\sigma }_{l}^{0,(n)}}}
\end{equation}
\bibliographystyle{IEEETran}
\bibliography{AE-Distarget}

\end{document}